\documentclass[reprint,amsmath,amssymb,aps,prb, superscriptaddress]{revtex4-1}

\usepackage{textcomp,pxfonts}
\usepackage{amsmath}
\usepackage{graphicx}
\usepackage{multirow}

\usepackage{color}
\begin{document}



\title{Advanced Post-Processing Techniques of Molecular Dynamics Simulations in Studying Strong Anharmonic Thermodynamics of Solids}

\author{Tian Lan} 
\email[]{tian@ginkgo-llc.com}
\affiliation{Department of Applied Physics and Materials Science, California Institute of Technology, Pasadena, California 91125, USA}
\affiliation{Ginkgo LLC, Incline Village, Nevada, 89452, USA}

\date{\today}

\begin{abstract}

While the vibrational thermodynamics of materials with small anharmonicity at low temperatures has been understood well based on the harmonic phonons approximation;
at high temperatures, this understanding must accommodate
how phonons interact with other phonons or with other excitations.
We shall see that the phonon-phonon interactions give rise to interesting coupling problems, and essentially modify the equilibrium and non-equilibrium properties
of materials, e.g., thermal expansion, thermodynamic stability, heat capacity, optical properties,
thermal transport and other nonlinear properties of materials. 
To date the anharmonic
lattice dynamics is poorly understood despite its great importance, and most studies on lattice dynamics
still rely on the harmonic or quasiharmonic models. 
With recent developement of computational models, the anharmonic information can be extracted from the atomic trajectories of molecular dynamics simulations. For example, the vibrational energy spectra, the effective potential energy surface and the phonon-phonon interaction channels can be derived from these trajectories which appear stochastic. These inter-dependent methods are adopted to successfully uncover the strong anharmonic phenomena while the traditional harmonic models fail dramatically, e.g., the negative thermal expansion of cuprite and the high temperature thermal stability of rutile. 

\end{abstract}


\maketitle


\section{Introduction}
Today our understanding of the vibrational thermodynamics of materials at low temperatures is emerging nicely, based on the harmonic model in which phonons are independent. At high temperatures, however, this understanding is generally poor since it must accommodate how phonons interact with other phonons (so called anharmonic phonon-phonon interactions ) or with electron or magnon excitations. These anharmonic processes and thermal excitations induce the frequency shifts and lifetime broadenings of the interacting quasi-particles, and contribute to most thermal properties at high temperatures.

These topics are rich and of great importance for the rational design and engineering of next-generation materials in energy based applications. For example, the anharmonic dynamics and other types of thermal excitations are the origin of most thermal energy transport processes, and therefore greatly influence the performance of these materials in applications of harvesting, storing and transporting energy. A reliable estimate of the anharmonic entropy is also crucial for synthesizing materials. For example, for metals and oxides, it seems that pure anharmonic contributions become large enough to affect phase stability at temperatures above half the melting temperature, which is the temperature range where materials are often processed or used. Anomaly in thermal expansion is another prominent example. Recently, the large negative thermal expansion (NTE) of ScF3 and Ag2O was found to have strong dependence with these high temperature vibrational dynamical properties. \cite{ScF3,Lan2014} 

Modern inelastic scattering techniques with neutrons or photons are ideal for sorting these properties out. Analysis of the experimental data can generate vibrational spectra of the materials, i.e., their phonon densities of states (DOS)and phonon or spin wave dispersions. We are developing the data reduction software to obtain the high quality data from inelastic neutron spectrometers. With accurate phonon DOS and dispersion curves we can obtain the vibrational entropies of different materials. The understanding of the underlying reasons for differences in DOS curves and entropies then relies on the development of the fundamental theories and the computational methods.

To date, most ab-initio methods for calculating materials structures and properties have been based on density functional (DFT) methods, and evaluating the internal energy, $E$, of materials at a temperature of zero Kelvin. For example, a harmonic or quasiharmonic model usually used to account for the vibrational thermodynamics at low temperatures and it is commonplace today to calculate harmonic phonons by methods based on DFT.\cite{Fultz2010} This can be adequate when the temperatures of service of the materials are low or when differences of chemical potential are much larger than kT. However, for most applications of materials in energy involving even modest temperature, this $E$ alone is insufficient because the anharmonic vibrational dynamics and different types of thermal excitations become to play important roles and have significant thermodynamic effects at elevated temperatures. 

Phonon-phonon interactions are responsible for 
pure anharmonicity that shortens phonon lifetimes and shifts phonon frequencies, especially at high temperatures. 
Anharmonicity competes with quasiharmonicity to alter the stability of phases at high temperatures, 
as has been shown, for example, with experiments and frozen phonon calculations 
on bcc Zr \cite{Ye1987} and
the possible stabilization of bcc Fe-Ni alloys at conditions of the Earth's core.\cite{Dubrovinsky2007}
For PbTe, ScF$_3$ and rutile TiO$_2$, there are recent reports of anharmonicity  being so large
that both the QHA and anharmonic perturbation theory 
fail dramatically.\cite{Delaire2011, ScF3, Lan2015}

These cases are suitable for ab initio
molecular dynamics (AIMD) simulations, 
which should be reliable when the electrons 
are near their ground states and the nuclear motions are classical. 
The big advantage of ab-initio MD is that it can account for all effects of harmonic, anharmonic and even some of the electron-phonon interactions. However, advanced post-processing methodologies are requied to extract concrete information from these simulations. In the few examples where comparisons have been made to ab-initio MD, agreement has been surprisingly good even for highly anharmonic materials. Today, by validating these calculated results with inelastic scattering experiments with facilities such as the Spallation Neutron Source for neutrons, we can obtain scientific details about phonon-phonon interactions, electron-phonon interactions and other excitations at elevated temperatures. 

In this article, we discuss several MD-based computational techniques available just recently which proved useful for assessing the  anharmonic vibrational thermodynamics of solids. The computational details are discussed in Section~\ref{computation:methods}, followed by a few examples shown in Section~\ref{} and ~\ref{} that demonstrate how the applications of these inter-dependent methods can uncover interesting anharmonic properties of materials and their relationships with NTE, vibrational energy shift and phase stability.   

\section{Computational Methodologies}
\label{computation:methods}

\subsection{Quasiharmonic Approximation}
The quasiharmonic approximation (QHA) is based on how phonon frequencies change with volume. 
In the QHA, all shifts of phonon
frequencies from their low temperature values are the result of thermal expansion alone.\cite{Fultz2010, Chen2015}
Although the QHA accounts for some frequency shifts, 
the phonon modes are still assumed to be harmonic, non-interacting, 
and their energies depend only on the volume
of the crystal.

In the quasiharmonic approximation, the vibrational free energy can be minimized 
as a function of volume,
\begin{eqnarray}
\! \! \! \! \! \! F(V,T) = E_0 + \int\limits_{- \infty}^{+ \infty} g( \omega ) \left(  \frac{\hbar \omega}{2}  + k_{\rm B }T \ln(1 - {\rm e}^{- \hbar \omega / k_{\rm B }T})  \right)
{\rm d} \omega \; ,
\end{eqnarray}
where $E_0 $ is the energy calculated for the relaxed structure at $T = 0 \,$K. Thermodynamic properties are therefore derived from here.\cite{Fultz2010,Chen2015}
  
\subsection{Molecular Dynamics Simulation}   
\subsubsection{Molecular Dynamics Simulation and Fourier transformed velocity autocorrelation method}
In non-harmonic potentials, phonon-phonon interactions are responsible for 
pure anharmonicity that shortens phonon lifetimes and shifts phonon frequencies. 
Anharmonicity competes with quasiharmonicity to alter the stability of phases at high temperatures.
Velocity trajectories were extracted from the MD simulation at each temperature, and were then transformed to the corresponding vibrational energy and/or momentum domain.\cite{Koker2009, MD2010, Lan2012, Lan2014, Lan2015, thesis}
Because the FTVAC method does not assume a form for the Hamiltonian,
it is a robust tool for obtaining  vibrational spectra from MD simulations, even with strong anharmonicity.
The phonon DOS is given by
\begin{equation}
\label{eq:DOScal}
g(\omegaup)=\sum_{n,b} \,\int e^{-{\rm i} \omegaup t}  \langle \vec{v}_{n,b}(t)\,\vec{v}_{0,0}(0) \rangle \, {\rm d} t \; ,
\end{equation}
where $\langle \, \rangle$ is an ensemble average, and $\vec{v}_{n,b}(t)$ is the velocity of the atom $b$ in the unit cell $n$ at time $t$.
Further projection of the phonon modes onto each $k$ point in the Brillouin zone was performed by computing the phonon power spectrum with the FTVAC method, with a resolution determined by the size of the supercell in the simulation.  

\subsubsection{Temperature-dependent effective potential method}

In general, 
the cubic phonon anharmonicity contributes  
to both the phonon energy shift and the 
lifetime broadening, whereas 
the quartic anharmonicity contributes only to the phonon energy shift.\cite{Maradudin, Lan20122}
To distinguish the roles of cubic and quartic anharmonicity, 
the temperature-dependent effective potential (TDEP) method \cite{Hellman3, Lan2015} was used.  
In the TDEP method, 
an effective model Hamiltonian is used to sample the potential energy surface,
not at the equilibrium positions of atoms,
but at the most probable positions for a given temperature in an MD simulation \cite{Hellman3}
\begin{eqnarray}
\label{eq:hamil}
H &=& U_0+\frac{1}{2}\sum_i m {\bf p}_i^2+\frac{1}{2}\sum_{ij\alpha\beta} \phi_{ij}^{\alpha\beta}u_i^{\alpha} u_j^{\beta}\\ \nonumber
&&+\frac{1}{3!}\sum_{ijk\alpha\beta\gamma} \psi_{ijk}^{\alpha\beta\gamma}u_i^{\alpha} u_j^{\beta} u_k^{\gamma} \; ,
\end{eqnarray}
where $\phi_{ij}$ and $\psi_{ijk}$ are  second- and third-order force constants, ${\bf p}$ is  momentum,
and $u_i^\alpha$ is the Cartesian component $\alpha$ of the displacement of atom $i$.  
 %
In the fitting, the ``effective'' harmonic force constants $\phi_{ij}$ are renormalized by
the quartic anharmonicity. 
The cubic anharmonicity, however, is largely accounted for by the third-order force constants $\psi_{ijk}$, 
and can be understood  in terms of the third-order phonon self-energy  
that causes linewidth broadening.\cite{Maradudin} 

The resulting Hamiltonian
was used to obtain the renormalized phonon dispersions (TDEP spectra) accounting for both the anharmonic shifts $\Delta$, and broadenings  $\Gamma$, 
of the mode $\vec{q}j$. These are derived from the real and imaginary 
parts of the cubic self-energies $\Sigma^{(3)}$, respectively.\cite{Maradudin}

\begin{eqnarray}
\Delta(\vec{q} j;\Omega)&=& -\frac{18}{\hbar^2}\sum_{{ \vec{q}_1}j_1} \sum_{{\vec{q}_2}j_2} \big\vert V(\vec{q}j;{ \vec{q}_1}j_1;{ \vec{q}_2}j_2) \big\vert^2 \Delta(\vec{q}_1+\vec{q}_2-\vec{q})\nonumber \\
  &&\times \,  {\wp} \Big[ \frac{n_1+n_2+1}{\Omega+\omegaup_1+\omegaup_2}
-\frac{n_1+n_2+1}{\Omega-\omegaup_1-\omegaup_2} \nonumber \\
  &&\mbox{} \quad +\frac{n_1-n_2}{\Omega-\omegaup_1+\omegaup_2}
-\frac{n_1-n_2}{\Omega+\omegaup_1-\omegaup_2} \Big] \label{shift3}\\
\Gamma(\vec{q}j;\Omega)&=& \frac{18\pi}{\hbar^2}\sum_{{ \vec{q}_1}j_1} \sum_{{\vec{q}_2}j_2} \big\vert V(\vec{q}j;{\vec{q}_1}j_1;{\vec{q}_2}j_2) \big\vert ^2 \Delta(\vec{q}_1+\vec{q}_2-\vec{q})\nonumber   \\
 &&\times \big[ (n_1+n_2+1) \, \delta(\Omega-\omegaup_1-\omegaup_2)\nonumber \\
&&\mbox{} \quad +2(n_1-n_2) \, \delta(\Omega+\omegaup_1-\omegaup_2) \big]    \; ,
\label{broadening3}
\end{eqnarray}
where $\Omega$ is the renormalized phonon frequency and $\wp$ denotes the Cauchy principal part. 
The $V(.)$'s are elements of the Fourier transformed third order force constants $\psi_{ijk}$ obtained in the TDEP method. The $\Delta(\vec{q}_1+\vec{q}_2-\vec{q})$ ensures conservation of momentum.

\subsection{Kinematics Functional of Phonon-Phonon Interactions}
Anharmonicity tensors describe the coupling strengths for phonon-phonon interactions, but
a prerequisite is that the phonons in these processes satisfy the kinematical conditions of 
conservation of energy and momentum as presented in Eq.~\ref{shift3} and~\ref{broadening3}.
In the phonon-phonon interaction functional, an anharmonicity tensor element 
for an $s$-phonon process can be expressed as \cite{Ipatova1967}
\begin{eqnarray}
\label{eq:tensor}
V(j;{\vec{q}_1}j_1;...;{\vec{q}_{s-1}}j_{s-1})&=&\frac{1}{2s!}\left(\frac{\hbar}{2N}\right)^{\frac{s}{2}}N \, \Delta({\vec{q}_1}+\cdots+{\vec{q}_{s-1}})\nonumber \\
\times[\omegaup_{j0}\omegaup_1 \cdots &\omegaup_{s-1}&]^{\frac{1}{2}}C({j;\vec{q}_1}j_1;...;{\vec{q}_{s-1}}j_{s-1})
\end{eqnarray} 
where $\Delta({\vec{q}_1}+\cdots+{\vec{q}_{s-1}})$ enforces 
momentum conservation and the $C(.)$'s, elements of the $s$-phonon anharmonic tensor,
are expected to be slowly-varying functions of their arguments. 

If the anharmonicity tensor or its average
does not vary significantly for
different phonon processes,
the coupling factor and the kinematic factor
are approximately separable in Eq.~\ref{eq:tensor}.
The separation of the anharmonic coupling and the kinematics
has been used with success in many studies including 
our recent reports on rutile TiO$_2$ and SnO$_2$. \cite{Lan2012, Lan20122} 
We consider the term $C(j;{\vec{q}_1}j_1;...;{\vec{q}_{s-1}}j_{s-1})$ 
to be a constant of the Raman mode $j$, and use it as a fitting parameter. 
Although
$C(j;{\vec{q}_1}j_1;{\vec{q}_{2}}j_{2})$ 
and $C(j;j;{\vec{q}_1}j_1;{-\vec{q}_1}j_1)$ change with
$j_1$ and $j_2$,  an average over modes, $\langle C(.) \rangle = \sum_{1,2} C(j;{\vec{q}_1}j_1;{\vec{q}_{2}}j_2)/ \sum_{1,2} 1$, is 
needed by the fitting, where 1, 2 under the summation symbol represent ${\vec{q}_i}j_i$.
We define the cubic and quartic fitting parameters as
\begin{subequations}
\label{eq:fitpara}
\begin{eqnarray}
C^{(3)}_j&=& \langle C(j;{\vec{q}_1}j_1;{\vec{q}_2}j_2) \rangle  \\
C^{(4)}_j&=& \langle C(j;j;{\vec{q}_1}j_1;{-\vec{q}_1}j_1) \rangle
\label{eq:fitpara2}
\end{eqnarray}
\end{subequations}

To the leading order of cubic and quartic anharmonicity, the broadening of the Raman peaks is  $2\Gamma^{(3)}(j;\Omega)$.
The frequency shift of the Raman peaks is $\Delta^Q+\Delta^{(3)}+\Delta^{(3^{\prime})}+\Delta^{(4)}$,
where the quasiharmonic part is denoted by $\Delta^Q$.
These quantities  can be written as  
functions of $D(\Omega,\omegaup_1,\omegaup_2)$ and $P(\Omega,\omegaup_1,\omegaup_2)$, weighted by average anharmonic coupling strengths\cite{Lan2012, Lan20122} 
\begin{subequations}
\label{eq:fitting}
\begin{eqnarray}
\! \! \! \! \! \! \! \! \! \! \! \! 
\Gamma^{(3)}(j;\Omega)&=&\frac{\pi\hbar}{64}\omegaup_{j0} \big\vert C^{(3)}_j \big\vert^2\sum_{{\vec{q}_1},j_1} \sum_{{\vec{q}_2},j_2}\omegaup_1\omegaup_2  \, D(\Omega,\omegaup_1,\omegaup_2) \nonumber\\
&=& \omegaup_{j0}\big\vert C^{(3)}_j \big\vert ^2 \, D^{\omegaup}(\Omega)\\
\! \! \! \! \! \! \! \! \! \! \! \!  
\Delta^{(3)}(j;\Omega)&=& -\frac{\hbar}{64}\omegaup_{j0} \big\vert C^{(3)}_j \big\vert ^2\sum_{{\vec{q}_1},j_1} \sum_{{\vec{q}_2},j_2}\omegaup_1\omegaup_2  \, P(\Omega,\omegaup_1,\omegaup_2)\nonumber\\
&=& \omegaup_{j0}\big\vert C^{(3)}_j \big\vert ^2 \, P^{\omegaup}(\Omega)\\
\! \! \! \! \! \! \! \! \! \! \! \!   
\Delta^{(3')}(j)&=&-\frac{\hbar}{16N}\omegaup_{j0} \big\vert C^{(3)}_j \big\vert^2 \sum_{{\vec{q}_2}j_2}\omegaup_{j_2}({\vec{q}_2}) \, \left( n_{{\vec{q}_2}j_2}+{\textstyle \frac{1}{2}} \right) \\
\! \! \! \! \! \! \! \! \! \! \! \!
\Delta^{(4)}(j)&=&\frac{\hbar}{8N}\omegaup_{j0}C^{(4)}_j\sum_{{\vec{q}_1}j_1}\omegaup_{j_1}({\vec{q}_1}) \, \left( n_{{\vec{q}_1}j_1}+{\textstyle \frac{1}{2}} \right)   
\end{eqnarray}
\end{subequations}
where $D^{\omegaup}(\Omega)$ and $P^{\omegaup}(\Omega)$ are  functionals of  $D(\Omega,\omegaup_1,\omegaup_2)$ and $P(\Omega,\omegaup_1,\omegaup_2)$ weighted by the kinematics of anharmonic phonon coupling. 
$D^{\omegaup}(\Omega)$ is the so-called two-phonon density of states (TDOS) spectra, which characterize the size of the phonon-phonon interaction channels.
The $\Delta^{(3^{\prime})}$ is an additional low-order cubic term that corresponds to  instantaneous three-phonon processes. \cite{Maradudin} 
It is nonzero for crystals having atoms without inversion symmetry, as in the case for the oxygen atom motions 
in the $A_{1g}$ mode of rutile. 
It is much smaller than other contributions, however, owing to symmetry restrictions.

\section{Negative Thermal Expansion of Cuprite Ag$_2$O and its Relationship with Strong Anharmonicity}
\label{nte:ag2o}

Silver oxide (Ag$_2$O) with the cuprite structure has attracted much interest after the discovery of its extraordinarily large negative thermal expansion (NTE),\cite{NTEa, NTEb} which exceeds $-1 \times 10^{-5}$ K$^{-1}$ 
and occurs over a wide range of temperature from 40\,K to its decomposition temperature near 500\,K.  

\begin{figure}[]
\includegraphics[width=0.8\columnwidth]{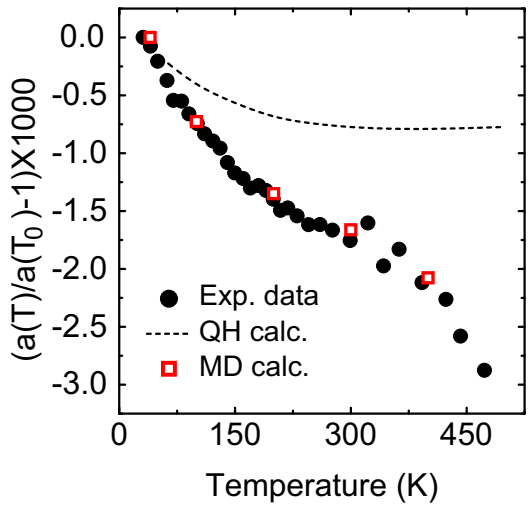}
\caption{Temperature dependence of  lattice parameter from experimental data in Ref. [\onlinecite{NTEa}], quasiharmonic calculations and MD calculations, expressed as the relative changes with respect to their 40\,K values, i.e., $a(T)/a(40\,K)-1$.}
\label{fig:Expansion}
\end{figure}

A rigid-unit modes (RUMs) model of NTE considers tetrahedra of Ag$_4$O around each O atom
that bend at the Ag atoms linking the O atoms in adjacent tetrahedra.
RUMs account for counteracting rotations of all such tetrahedra. \cite{RUMa, RUMb}
Locally, the O-Ag bond length does not contract, but bending of the 
O-Ag-O links pulls the O atoms together,
leading to NTE. These RUMs tend to have low frequencies owing to the large mass of the unit, 
and hence are excited at low temperatures.
This model correlateds the NTE with quasiharmonic approximation and should explain the main physics at low temperatures.
However, as shown in Fig.~\ref{fig:Expansion}, at temperatures above 250\,K, there is a second part of the 
NTE behavior of cuprite Ag$_2$O that is apparently beyond the predictions of  quasiharmonic theory.

\begin{figure*}[]
\includegraphics[width=1.8\columnwidth]{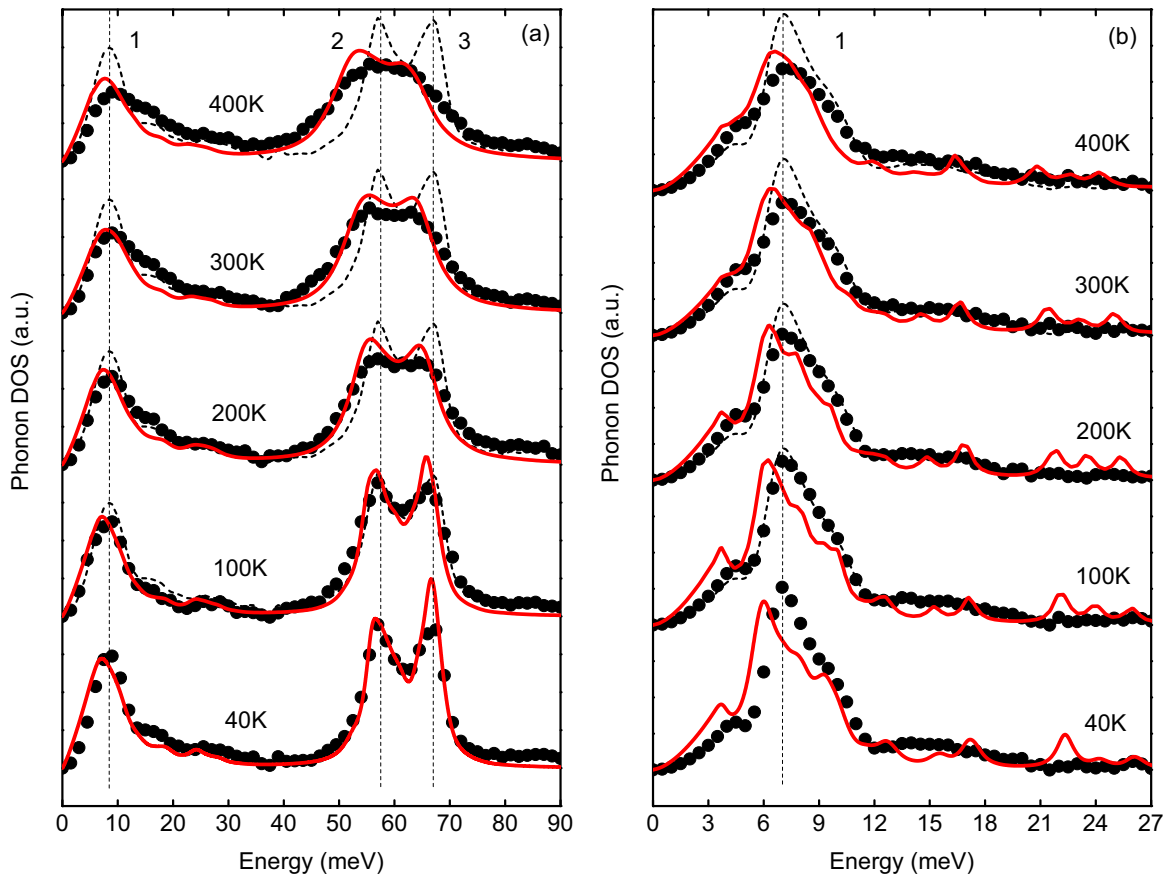}
\caption{Neutron weighted phonon DOS of  Ag$_2$O with the cuprite structure from ARCS experimental data (black dots) and MD simulations (red curves) at temperatures from 40 to 400\,K. The dashed spectrum corresponds to the 40\,K experimental result, 
shifted vertically for comparison at each temperature.  
Vertical dashed lines are aligned to the major peak centers at 40\,K from experiments, and are numbered at top. 
The incident energy was 100\,meV for panel (a), and 30\,meV for panel (b). 
}
\label{fig:Ag2ODOS}
\end{figure*}
 
\subsection{Computational Methods and Results}
First-principles calculations were performed with the generalized gradient approximation (GGA) of  density functional theory (DFT), implemented in the VASP package.\cite{vaspa,vaspb, vaspc}
Projector augmented wave pseudopotentials and a plane wave basis set with an energy cutoff of 500 eV were used in all calculations.  
 
First-principles Born-Oppenheimer molecular dynamics simulations were performed for a $3 \times 3 \times 3$ supercell with temperature  control by a Nos{\'{e}} thermostat. The relatively small simulation cell could be a cause for concern,\cite{sizeeffect} 
but  convergence testing showed that the supercell in our study is large enough to accurately capture the phonon anharmonicity of Ag$_2$O. The simulated temperatures included 40, 100, 200, 300 and 400\,K. 
For each temperature, the system was first equilibrated for 3 ps, then simulated for 18 ps with a time step of 3\,fs. The system was fully relaxed at each temperature, with convergence of the pressure within 1 kbar.

The phonon DOS curves calculated from first-principles MD simulations 
are shown in Fig.~\ref{fig:Ag2ODOS}
with the experimental spectra for comparison. 
Excellent agreement is found between the simulated phonon DOS and the experimental data,
and the calculated thermal broadenings and shifts are in good agreement, too. 

Because of the large mass difference between Ag and O atoms, 
the O-dominated phonon modes are well separated from the Ag-dominated modes. 
Partial phonon DOS analysis showed that
the Ag-dominated modes had similar energies, forming the peak of the phonon DOS below 20\,meV (peak 1 in Fig.~\ref{fig:Ag2ODOS}), whereas the O-dominated modes had energies above 40\,meV (peaks 2 and 3).

This NTE above 250\,K is predicted accurately by the ab-initio MD calculations,
so it is evidently a consequence of phonon anharmonicity. 
The temperature-dependence of this NTE behavior follows the Planck 
occupancy factor for phonon modes above 50\,meV, corresponding to the O-dominated band of optical frequencies.
In the QHA these modes above 50\,meV do not contribute to the NTE.
These modes are highly anharmonic, however, as shown by their large broadenings and shifts. 

\begin{figure}[t]
\includegraphics[width=1.0\columnwidth]{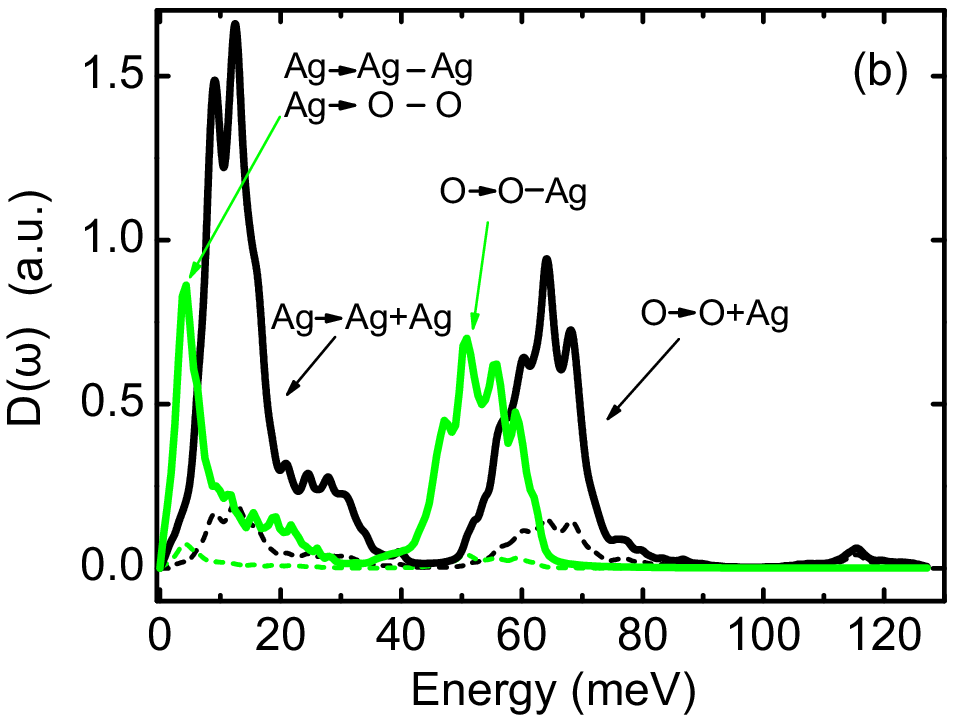}
\caption{
The TDOS spectra, $D(\omegaup)$, at 40\,K (dashed) and 400\,K (solid). The down-conversion and up-conversion contributions are  presented separately as  black and green curves, respectively. }
\label{fig:2DOS}
\end{figure} 

\subsection{Discussion} 
For  cubic anharmonicity, as discussed, the two-phonon DOS (TDOS) is the spectral quantity parameterizing the 
number of phonon-phonon interaction channels available to a phonon. 
For  Ag$_2$O with the cuprite structure, the peaks in the TDOS overlap well with the peaks in the phonon DOS. 
Most of the phonons therefore have many possible interactions with other phonons, which contributes to the large anharmonicity of 
 Ag$_2$O with the cuprite structure, and small ${\mathcal Q}$  (short lifetimes). 
Although the ${\mathcal Q}$ values of most phonon modes in  Ag$_2$O with the cuprite structure
are small and similar, the origins of these lifetime broadenings are intrinsically different. 
For  peak 2 of the phonon DOS, the anharmonicity is largely from the up-conversion 
processes: O $\mapsto$ O $-$ Ag, while for peak 3 it is from the down-conversion 
processes: O $\mapsto$ O $+$ Ag. 
The anharmonicity of peak 1 is more complicated. 
It involves both  up-conversion and down-conversion processes of Ag-dominated modes. 
The TDOS also shows why the  $A_{2u}$ mode has a larger ${\mathcal Q}$  than other modes. 
Figure~\ref{fig:2DOS}(b) shows that the  $A_{2u}$ mode lies in the trough of the TDOS 
where there are only a few  phonon decay channels. 

Owing to explicit anharmonicity from phonon-phonon interactions, the thermodynamic 
properties of  Ag$_2$O with the cuprite structure cannot be understood as a sum of contributions from independent normal modes. 
The frequency of an anharmonic phonon depends on the level of excitation of other modes. 
At high temperatures, large vibrational amplitudes increase the anharmonic coupling of modes, 
and this increases the correlations between the motions of the Ag and O atoms, as shown by perturbation theory. 
Couplings in perturbation theory have phase coherence, 
so the coupling between Ag- and O-dominated modes at higher energies,
as seen in the peak of the TDOS,
causes correlations
between the motions of Ag and O atoms. 
The ab-initio MD simulations
show that anharmonic interactions allow 
the structure to become more compact with increasing vibrational amplitude. 
The mutual motions of the O and Ag atoms cause higher density as the atoms fill space more effectively.
The large difference in atomic radii of Ag and O may contribute to this effect.
Perhaps it also facilitates the irreversible changes in Ag$_2$O
at temperatures above 500\,K, but this requires further investigation. 
For cuprite Cu$_2$O, which has less of a difference in atomic radii, 
the thermal expansion is much less anomalous.

\section{The Quartic Phonons and its Stabilization of Rutile Phase of TiO$_2$}
\label{quarticity:tio2}
Although the rutile structure of TiO$_2$ is stable at high temperatures,
the conventional quasiharmonic approximation predicts that several  acoustic
phonons decrease anomalously to zero frequency with thermal expansion,
incorrectly predicting a structural collapse at temperatures well below 1000\,K.\cite{Mitev2010, Refson2013}
Inelastic neutron scattering  was used to measure the 
temperature dependence of the phonon density of states (DOS) of rutile TiO$_2$ from 300 to 1373\,K.
Surprisingly, these anomalous acoustic phonons were found to increase in frequency with temperature.\cite{Lan2015}

\subsection{Computational Methods and Results}
First-principles calculations using the local density approximation (LDA) of  density functional theory (DFT) were performed with the VASP package.\cite{vaspa, vaspc}
Projector augmented wave pseudopotentials and a plane wave basis set with an energy cutoff of 500 eV were used in all calculations. 
Previous work showed that for best accuracy, the Ti pseudopotential should treat the semicore 3$s$ and 3$p$ states as valence,\cite{Harrison2004,  Mitev2010, Refson2013}
and we took this approach with a similar LDA functional. 
Our calculated  elastic properties, lattice dynamics, and dielectric properties derived from the optimized structure for 0\,K, were in good agreement with results from experiment and from previous DFT calculations.

First-principles Born-Oppenheimer AIMD simulations for a $2 \times 2 \times 4$ supercell and a $2 \times 2 \times 1$ $k$-point sampling were performed to thermally excite phonons to the target temperatures of 300 and 1373\,K. 
For each temperature, the system was first equilibrated for 3\,ps as an $NVT$ ensemble with temperature control by a Nos{\'{e}} thermostat, then simulated as an $NVE$ ensemble for 20\,ps with time steps of 1\,fs. 
Good relaxations with residual pressures below 0.5\,GPa were achieved in each calculation that accounted for  thermal expansion.

\begin{figure*}[t]
\includegraphics[width=2.1\columnwidth]{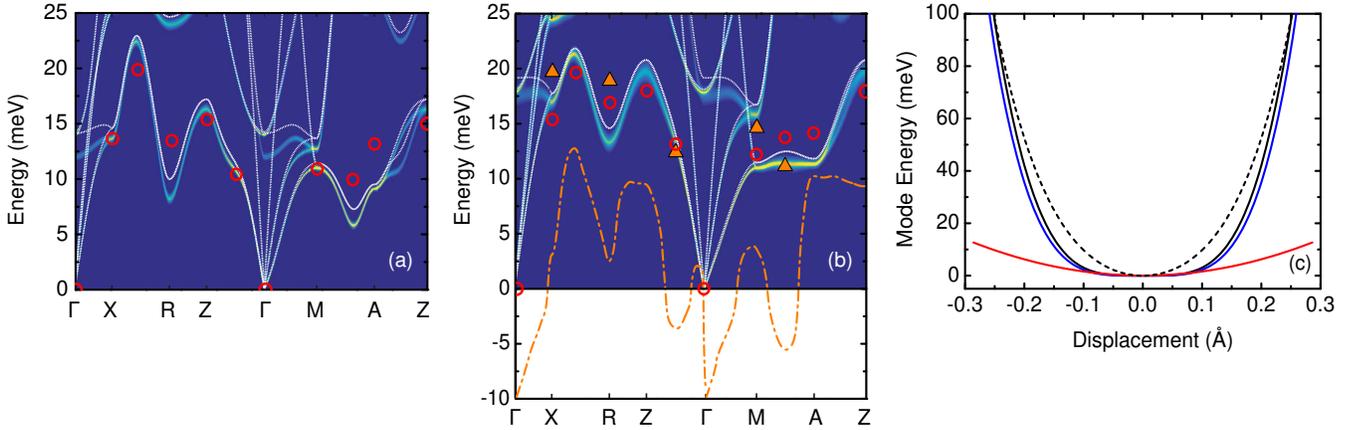}
\caption{Diffuse curves are 
TDEP phonon dispersions 
below 25\,meV at (a) 300\,K, and (b) 1373\,K,  compared with the results from the FTVAC method (red circles). The white curves are phonon dispersions for the quasiharmonicity plus quartic anharmonicity calculated with all $\psi_{ijk}$ set to zero in Eq.~\ref{eq:hamil}.  In (b), the dispersions are compared to the quasiharmonic dispersions (orange dashed curve) and the  single quartic oscillator model (orange triangles).  
(c) Frozen phonon potential (black) of TA mode  at R point with $q = (0.5, 0, 0.5)$ at 1373\,K,
showing the harmonic component (red) and quartic component (blue). The 
low temperature potential surface is also shown (dashed black). 
}
\label{fig:quartic}
\end{figure*}

Figure~\ref{fig:quartic} shows the 
vibrational energies of  the TA branch,
calculated by the FTVAC method with AIMD trajectories. 
From 300 to 1373 \,K, the TA branch increases in energy by an average 
of about 2.1\,meV. 
Especially for this TA branch, 
Fig.~\ref{fig:quartic}(b) shows an enormous 
discrepancy of phonon energies between the MD calculation and the QHA (orange dashed line) at 1373\,K. 
The unstable phonon modes predicted by the QHA  are fully stable   
in the AIMD simulations at high temperatures, however.

Using the same MD trajectories as for the FTVAC method, 
the calculated TDEP dispersions agree well with the FTVAC results as shown in Fig.~\ref{fig:quartic}(a)(b).
The cubic anharmonicity of rutile TiO$_2$ 
is strong for phonons at energies above 25\,meV, \cite{Lan2012}
causing broadening of the phonon DOS and the dispersions.
Nevertheless, at 1373\,K the TA modes below 20\,meV have 
only small linewidth broadenings.
Furthermore, they  are  close in energy to those calculated 
if all $\psi_{ijk}$ are set to zero in 
Eq. \ref{eq:hamil}, showing the dominance of quartic anharmonicity and the
small cubic anharmonicity of the TA modes.

For more details about  the anomalous anharmonicity of the TA modes, 
the frozen phonon method was used to calculate potential energy surfaces for specific phonons, as in
Fig.~\ref{fig:quartic}(c). 
At 300\,K the potential energy of the TA mode at the $R$-point is nearly quadratic,
with a small quartic part. 
With the lattice expansion characteristic of 1373\,K, 
the potential energy curve transforms to being nearly quartic. 
In fact, for all modes in the TA branch that were evaluated by the frozen phonon method, 
the potential energy surface develops a quartic form with lattice expansion
For a quantum quartic oscillator, the vibrational frequency stiffens with temperature
owing to the increasing spread between the energy levels.\cite{Dorey1999,ScF3}
We  assessed a high temperature behavior 
by assigning
Boltzmann factors to the different oscillator
levels derived from frozen phonon potentials, giving 
the energies of the quartic TA modes at 1373\,K. 
As shown in Fig.~\ref{fig:quartic}(b),  they are reasonably close to the FTVAC and TDEP results.

\begin{figure*}[]
\includegraphics[width=1.8\columnwidth]{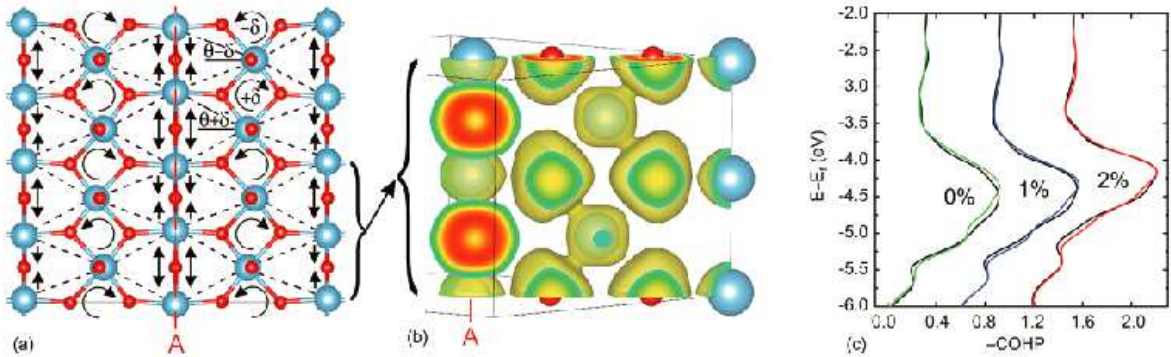}
\caption{(a) Displacements of atoms for the TA mode at the R-point in the (1-10) plane. Light blue spheres are Ti atoms, and O atoms are red. Arrows depict distortions of the structural units (dashed rhombuses). The rotational movements of structures, or the ``ring" patterns, are indicated with circled arrows. 
(b) ELF isosurface of a ``ring" shown in (a) with the value of 0.3. The ELF increase is apparent in the bond of shorter distance owing to the ring displacement. 
The ELF is the probability measure of finding an electron at a
location given the existence of neighbouring electrons.
It ranges from 0 (no electron) to 1 (perfect localization).  
(c) COHP analysis of Ti--O bonds for equilibrium lattice parameter at $T$=0\,K (0\%), and for linear expansions of 1\% and 2\%. Shown in color are --COHP results for the same structures with the phonon of panel (a) having 0.14\,${\rm \AA}$ normal displacements of Ti atoms. 
Curves for 1\% and 2\% expansion are offset  by 0.6 and 1.2. }
\label{fig:pattern}
\end{figure*}

\subsection{Discussion}
The  patterns of atom displacements 
in the anomalous modes at $\Gamma$, R, and along Z-$\Gamma$, $\Gamma$-M, and M-A
were identified, and  those for the R point in Fig. \ref{fig:pattern}(a) are typical.
In these anomalous modes, the O atoms were approximately stationary,
and each O atom has a Ti neighbor that moves towards it and another Ti neighbor that moves away from it by approximately the same amount.
These modes have ``ring'' patterns in which displacements of Ti atoms rotate  
a structural unit, and all the O atoms see approximately the same change in their Ti neighborhood.
In the positive and negative displacements of these modes,
the O atoms show
an accumulation of charge in the Ti--O bond of shorter distance
and a depletion in the bond of longer distance, as indicated by a much higher value of the electron localization function (ELF) \cite{ELF} for the short bond 
shown in Fig.~\ref{fig:pattern}(b).

We calculated the ``bond-weighted'' electron DOS by partitioning the band structure energy into bonding and no-bonding contributions and obtaining the crystal orbital Hamilton population (COHP) spectrum \cite{cohp2} of rutile with different lattice parameters in a $2 \times 2 \times 4$ supercell.  Figure \ref{fig:pattern}(c) shows the COHP spectrum of the 
bonds formed by the Ti\,$3d$\, and O\,$2p$ orbitals between
5.5 and 3.5\,eV below the Fermi energy. 
With lattice expansion (of 1\% or 2\%), these bonding states become less favorable, and their COHP spread
narrows. 
Also shown in color in Fig. \ref{fig:pattern}(b) is the COHP with a frozen phonon mode at the R-point
having 0.14\,${\rm \AA}$ displacements of  Ti atoms. 
On the scale of thermal energies, the broadening effect from the  phonon changes considerably with lattice expansion. 

With an increase in lattice parameter,
the longer Ti--O bond makes a smaller contribution to the interatomic force during its vibrational cycle.
The shorter Ti--O bond gives a stronger hybridization of  Ti\,$3d$\, -- O\,$2p$ orbitals as the Ti atom moves closer to the O atom.
The hybridization serves to offset the energy of short-range repulsion. 
With lattice expansion, the short-range repulsion is weaker, and 
hybridization favors electrons  
between the shorter Ti--O pairs in the phonon displacement pattern. 
The ``ring'' patterns of the phonons play an important role in increasing the degree of hybridization as they 
complete the electron back-donation cycles from the O to the Ti atoms. For example, a 3\% decrease of Bader effective charge (+2.22 at equilibrium) was found for the Ti atoms with 0.14 \,${\rm \AA}$ displacments in the ``ring" patterns, which is comparable to the charge decrease of the Ti atoms during the ferroelectric transition of rutile.\cite{Harrison2004} However,
if Ti atoms along the direction ``A'' were locked down at their 
equilibrium positions so the ring motion is broken, the resulting decrease of the effective charge dropped  by 50\% to 70\% in the ``ring" patterns. The potential 
was found to rise, and was largely quadratic even at 1\% or 2\% expansion.

A macroscopic elastic response to this phonon can also be identified with the assistance of Fig. \ref{fig:pattern}(a). 
In equilibrium, the apex angles of the rhombuses are all equal, but with the rotation by $\delta$, the vertical stretching of  rhombuses along the line A is $ 2a \sin ( \theta + \delta )$, and the contraction is $ 2a \sin ( \theta - \delta )$, where $\theta$ is the semi-angle of the rhombus. For small $\delta$, a Taylor expansion gives a net vertical (or horizontal) distortion of $  - 2a \delta^2 \sin \theta $ (or $  - 2a \delta^2 \cos \theta$). The distortions are proportional to $\delta ^2$, while the atom displacements in this TA mode are proportional to $\delta $. 
A strain energy that goes as the square of this distortion is consistent with a quartic potential. 

The hybridization in the Ti--O bond  is very sensitive to interatomic distance, much as has been noticed 
in the ferroelectric distortion of BaTiO$_3$.\cite{ferro1}
For rutile TiO$_2$, however, the 
hybridization follows the atom displacements in thermal phonons (instead of a displacive phase transition),
and  this  ``phonon-tracked hybridization''  changes with lattice parameter. 
It provides a source of extreme phonon anharmonicity, but also provides thermodynamic stability for  rutile TiO$_2$.
It may occur in other transition metal oxides that show unusual changes of properties with lattice parameter or with structure, and such materials may be tunable with composition or pressure to control this effect. Besides altering thermodynamic phase stability, properties such as ferroelectricity and thermal transport will be affected directly. 

\section{Conclusions}
\parskip=0.in
The study of phonon anharmonicity and phonon-phonon interaction is a difficult
but exciting field. It is difficult because we must consider how phonons interact
with other phonons or with other excitations, which is an example of notorious manybody
interaction problem. In comparison, our understanding today about the
vibrational thermodynamics of materials at low temperatures is broad and deep
because it is based on the harmonic model in which phonons are independent,
avoiding issues of anharmonic lattice dynamics. However, the failure of the harmonic theory also mostly arises from the assumption of independent
phonons, which becomes increasingly inaccurate at high temperatures. 

With the development of molecular dynamics simulations and the progress of the anharmonic phonon theories and computational
methodologies based upon it, we are now in a good position to study the relation between the
phonon anharmonicity and many important thermodynamic properties of materials especially at high temperatures.
For example, advanced computational methods being discussed in this paper provided the microscopic perspective of phonon anharmonicty and its relationship with the thermodynamic phase stability of rutile TiO$_2$ and anomaly thermal expansion of cuprite Ag$_2$O.

\bibliography{ref}

\newpage




\end{document}